\documentclass[preprint,aps,showpacs]{revtex4}
\usepackage{graphicx}
\usepackage{epstopdf}
\usepackage{dcolumn}
\usepackage{bm}

\begin{document}


\title{Towards an ideal graphene-based spin-filter}

\author{Yu.\,S.\,Dedkov,$^{1,2,}$\footnote{Corresponding author. E-mail: dedkov@fhi-berlin.mpg.de} A. Generalov,$^{1,2}$ E.\,N.\,Voloshina,$^{3}$ and M. Fonin$^{4}$}
\affiliation{$^1$Fritz-Haber-Institut der Max-Planck-Gesellschaft, 14195 Berlin, Germany}
\affiliation{$^2$Institut f\"ur Festk\"orperphysik, Technische Universit\"at Dresden, 01062 Dresden, Germany}
\affiliation{$^3$Institut f\"ur Chemie und Biochemie-Physikalische und Theoretische Chemie, Freie Universit\"at Berlin, Takustrasse 3, 14195 Berlin, Germany}
\affiliation{$^4$Fachbereich Physik, Universit\"at Konstanz, 78457 Konstanz, Germany}

\date{\today}

\begin{abstract}
The Fe$_3$O$_4$(111)/graphene/Ni(111) trilayer is proposed to be used as an ideal spin-filtering sandwich where the half-metallic properties of magnetite are used. Thin magnetite layers on graphene/Ni(111) were prepared via successive oxidation of thin iron layer predeposited on graphene/Ni(111) and formed system was investigated by means of low-energy electron diffraction (LEED) and photoelectron spectroscopy (PES) of core levels as well as valence band. The electronic structure and structural quality of the graphene film sandwiched between two ferromagnetic layers remains unchanged upon magnetite formation as confirmed by LEED and PES data.  
\end{abstract}

\pacs{73.22.Pr, 79.60.-i, 61.05.jh}

\maketitle

\begin{figure}[t]
\includegraphics[scale=1.5]{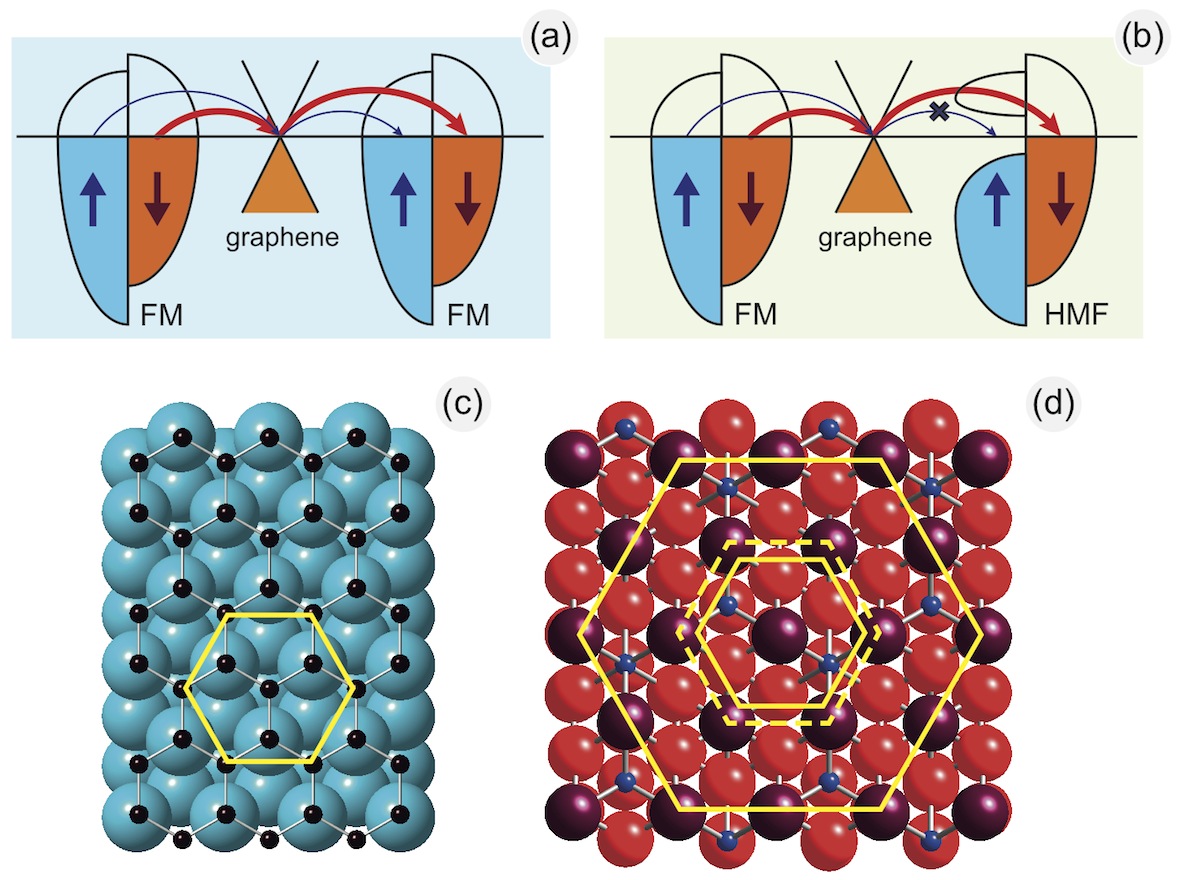}
\newline
\newline
\caption{(Color online).}
\label{schemeANDstr}
\end{figure}

Graphene~\cite{Geim:2007a,Geim:2009}, a two-dimensional honeycomb lattice-packed layer of carbon atoms, was recently proposed to be used as an effective spin-filtering material when sandwiched between two layers of ferromagnetic materials [Fig.~\ref{schemeANDstr}(a)]~\cite{Karpan:2007,Karpan:2008,Yazyev:2009}. For this purposes, the close-packed surfaces Ni(111)~\cite{Dedkov:2001,Dedkov:2008a,Dedkov:2010a}, Co(0001)~\cite{Eom:2009,Varykhalov:2009} or Fe(111)/Ni(111) interface~\cite{Dedkov:2008b,Weser:2011} can be used as a ferromagnetic substrate for the graphene preparation due to the extremely small mismatch between lattice constants of the graphene plane and substrates [see Fig.~\ref{schemeANDstr}(c)]. This spin-filtering effect is based on the fact that at the Fermi level the electronic structures of graphene and ferromagnetic close-packed (111) layers overlap only for one kind of spin~\cite{Karpan:2007}.

Several shortcomings hamper the realization of such graphene-based spin-filtering device. As was shown in Refs.~\cite{Karpan:2007,Karpan:2008} the use of only one layer of graphene in such device gives the magnetoresistive (MR) effect of approximately 20\% due to the strong chemical interaction between graphene and ferromagnetic (FM) layer [both spin-channels participate in the transport through the FM/graphene/FM trilayer as shown in Fig.~\ref{schemeANDstr}(a)]. It was demonstrated that MR ratio is drastically increased when multilayers of graphene are used. However, technologically the controllable growth of several layers of graphene is difficult due to the fact that graphene/FM system is very inert~\cite{Dedkov:2008} and preparation of the second and the third layer on it requires long time exposure of this system to hydrocarbons at high temperature~\cite{Nagashima:1994a}. Preparation of the FM/graphene/FM sandwiches via the so-called ``scotch-tape'' method~\cite{Hill:2006,Geim:2007a} or application of the segregation-from-bulk method~\cite{Yu:2008,Li:2009a}, where graphene multilayers can be easily produced on top of ferromagnetic material, can not be considered as appropriate approaches because in both cases the interface properties between graphene and FM or bulk properties of FM material are strongly affected during the preparation procedure and can not be easily modified in a controllable way.

One more aspect, which can make the realization of the graphene-based spin-filter in the current-perpendicular-to-plane (CPP) geometry difficult, is the reduction of the stability of graphene upon top-ferromagnetic contact deposition. First of all, ferromagnetic metals (Fe, Co, Ni) deposited on top of the graphene/metal interface have a trend to form clusters~\cite{NDiaye:2009a,Sicot:2010,Lahiri:2010a} without preferential crystallographic orientation, that destroys the condition when the $K$ points of the Brillouin zones of a graphene lattice and the close-packed surface of FM are coincide in the reciprocal space~\cite{Karpan:2007,Karpan:2008}. Moreover, the strong chemical interaction between graphene and the top ferromagnetic electrode reduces the stability of the graphene layer~\cite{Lahiri:2010a} that limits the thermal treating of the FM/graphene/FM sandwich during preparation.

Here we propose an application of structurally and electronically stable thin layer of half-metallic ferromagnetic (HMF) material, magnetite (Fe$_3$O$_4$), as a second top-ferromagnetic electrode in the HMF/graphene/FM spin-filter (HMF is a ferromagnetic material which has only one kind of spin at the Fermi level) [Fig.~\ref{schemeANDstr}(b)]. This configuration might help to overcome major technical problems discussed above. First of all, the perfect crystallographic matching is not necessary in this case, because density of states of HMF Fe$_3$O$_4$(111) layer [see Fig.~\ref{schemeANDstr}(d)] is fully spin-polarized at the Fermi level~\cite{Dedkov:2002a,Zhou:2010} and only one kind of spin can be detected in HMF/graphene/FM trilayer. Secondly, the electronic interaction between graphene and a magnetite layer supposed to be extremely small due to the chemical inertness of both materials. In our work we demonstrate that the Fe$_3$O$_4$(111)/graphene/Ni(111) trilayer can be prepared in a controllable way. All preparation steps and crystallographic as well as electronic structure were controlled by means of low-energy electron diffraction (LEED) and photoelectron spectroscopy (PES) of core levels as well as valence band. Thin magnetite layers were prepared via successive oxidation of thin iron layer predeposited on graphene/Ni(111). It is found that the electronic structure and structural quality of the graphene layer sandwiched between two ferromagnetic layers in such trilayer remains unchanged upon magnetite formation as confirmed by diffraction and photoelectron spectroscopy data.

The present studies were performed in an experimental setup for photoelectron spectroscopy consisting of two chambers described in detail elsewhere~\cite{Dedkov:2008,Dedkov:2008a,Dedkov:2008b}. As a substrate a W(110) single crystal was used. Prior to preparation of the studied systems a well-established cleaning procedure of the W-substrate was applied. A well ordered Ni(111) surface was prepared by thermal deposition of Ni films with a thickness of about 200\,\AA\ on a clean W(110) substrate and subsequent annealing at $300^\circ$C. An ordered graphene overlayer was prepared via cracking propene gas (C$_3$H$_6$) according to the recipe described in Ref.~\cite{Dedkov:2008,Dedkov:2008a,Dedkov:2008b}. After the cracking procedure the Ni(111) surface is completely covered by the graphene film as was earlier demonstrated in Refs.~\cite{Dedkov:2008a,Dedkov:2008b}. Preparation of Fe$_3$O$_4$ on top of the graphene/Ni(111) system was performed via subsequent oxidation of thin predeposited Fe film followed by annealing at a moderate temperature. Photoemission spectra monitoring the process of system preparation were recorded at 21.2\,eV, 40.8\,eV (He\,I$\alpha$, He\,II$\alpha$, UPS) as well as 1253.6\,eV, 1486.6\,eV (Mg\,K$\alpha$, Al\,K$\alpha$, XPS) photon energies using a hemispherical energy analyzer SPECS PHOIBOS 150. The energy resolution of the analyzer was set to 50\,meV and 500\,meV for UPS and XPS, respectively.

\begin{figure}[t]
\includegraphics[scale=1.0]{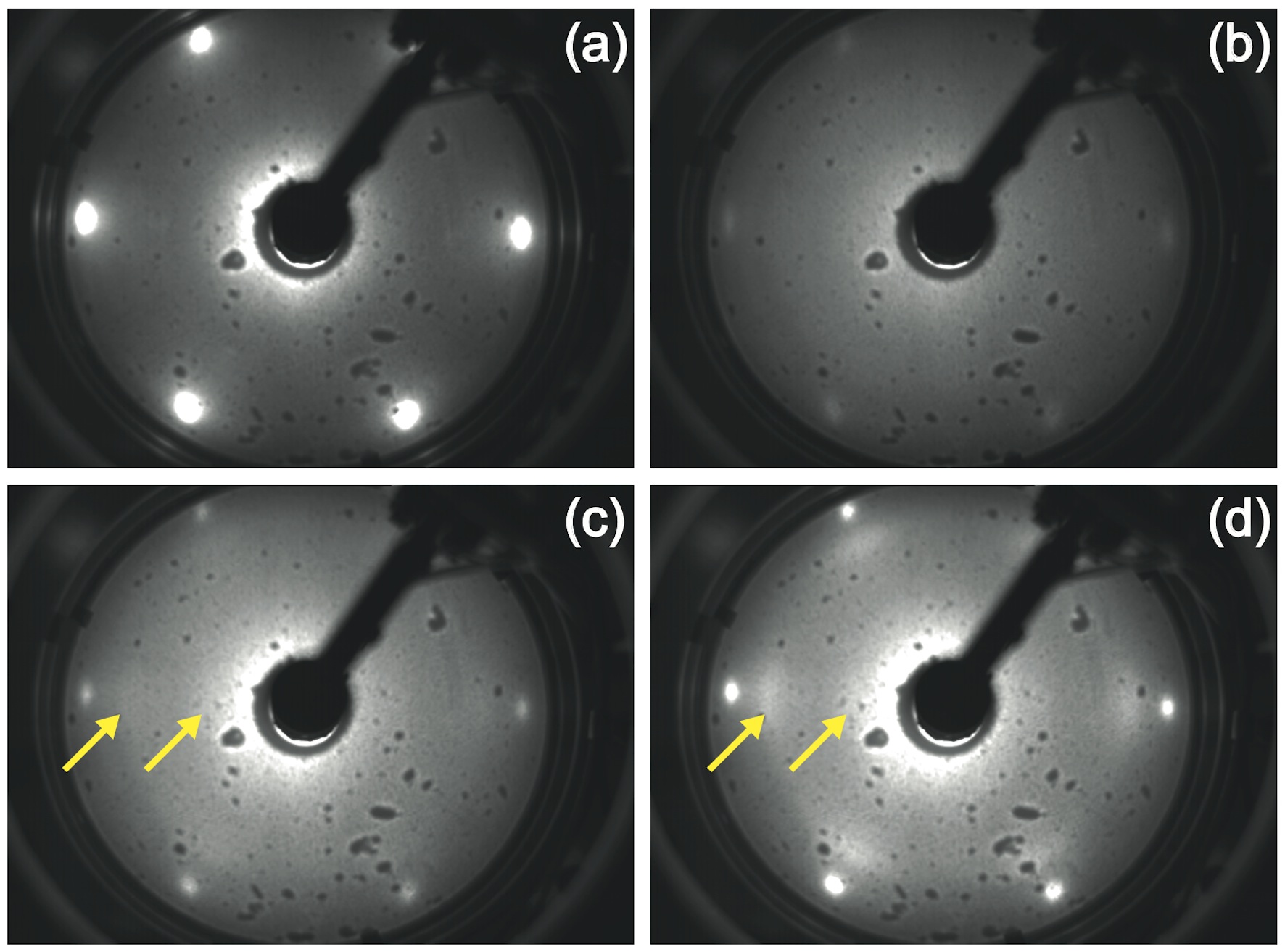}
\newline
\newline
\caption{(Color online).}
\label{LEED}
\end{figure}

The formation of the Fe$_3$O$_4$(111)/graphene/Ni(111) trilayer was simultaneously monitored by LEED and PES of core levels as well as valence band of the system. Fig.~\ref{LEED}(a) shows the LEED image of the graphene/Ni(111) system prepared on the W(110) substrate which reveal a well-ordered	 $p(1\times1)$ overstructure as expected from the small lattice mismatch of only 1.3\% between graphene plane and Ni(111). Deposition of 10\,\AA\ of Fe on top of this system leads to a weakening of the LEED picture from graphene/Ni(111) (weak diffraction spots can still be recognized) with diffuse background in the image (LEED is not shown). Exposure of the 10\,\AA\,Fe/graphene/Ni(111) system to large amount of oxygen [$p$(O$_2$)$=1\times10^{-6}$\,mbar for 1\,hour] leads to very diffuse background with extremely weak spots from graphene/Ni(111) [Fig.~\ref{LEED}(b)]. In order to improve the crystallinity of the system after oxidation of the top Fe layer, it was annealed at $T=150^\circ$C for 3 min. This treatment leads to the changes in the LEED picture [Fig.~\ref{LEED}(c)], namely: (i) diffraction spots from graphene/Ni(111) become slightly sharper that can be explained by smoothening of the top iron oxide layer and (ii) additional blurred ring-shaped reflexes appear in LEED which are marked by light arrows in Fig.~\ref{LEED}(c). Here, the inner and outer spots can be considered as a first- and second-order diffraction spots from the same structure. Further annealing of the system at higher temperature $T=250^\circ$C for the same time leads to the picture shown in Fig.~\ref{LEED}(d): sharp diffraction spots from the graphene/Ni(111) system and easily-recognizable additional ring-shaped spots. On the basis of XPS and PES analysis (see discussion below in the text) one can conclude that sharp Fe$_3$O$_4$/graphene interface is formed in both annealing cases. Hexagonal symmetry of LEED spots shown in Fig.~\ref{LEED}(c,d) and the absence of any moir\'e-like superstructures in the image indicate the formation of the Fe$_3$O$_4$(111) thin film which does not interact electronically with the underlying graphene layer on Ni(111) (also supported by XPS and PES results; see discussion below). The estimated from LEED lattice parameter of the Fe$_3$O$_4$(111) surface [light big hexagon in Fig.~\ref{schemeANDstr}(d)] is $6.02\pm0.15$\,\AA\ which is very close to values obtained in the previous experiments: $5.92-6.0$\,\AA~\cite{Weiss:1993,Shaikhutdinov:1999,Dedkov:2004a}.

\begin{figure}[t]
\includegraphics[scale=1.0]{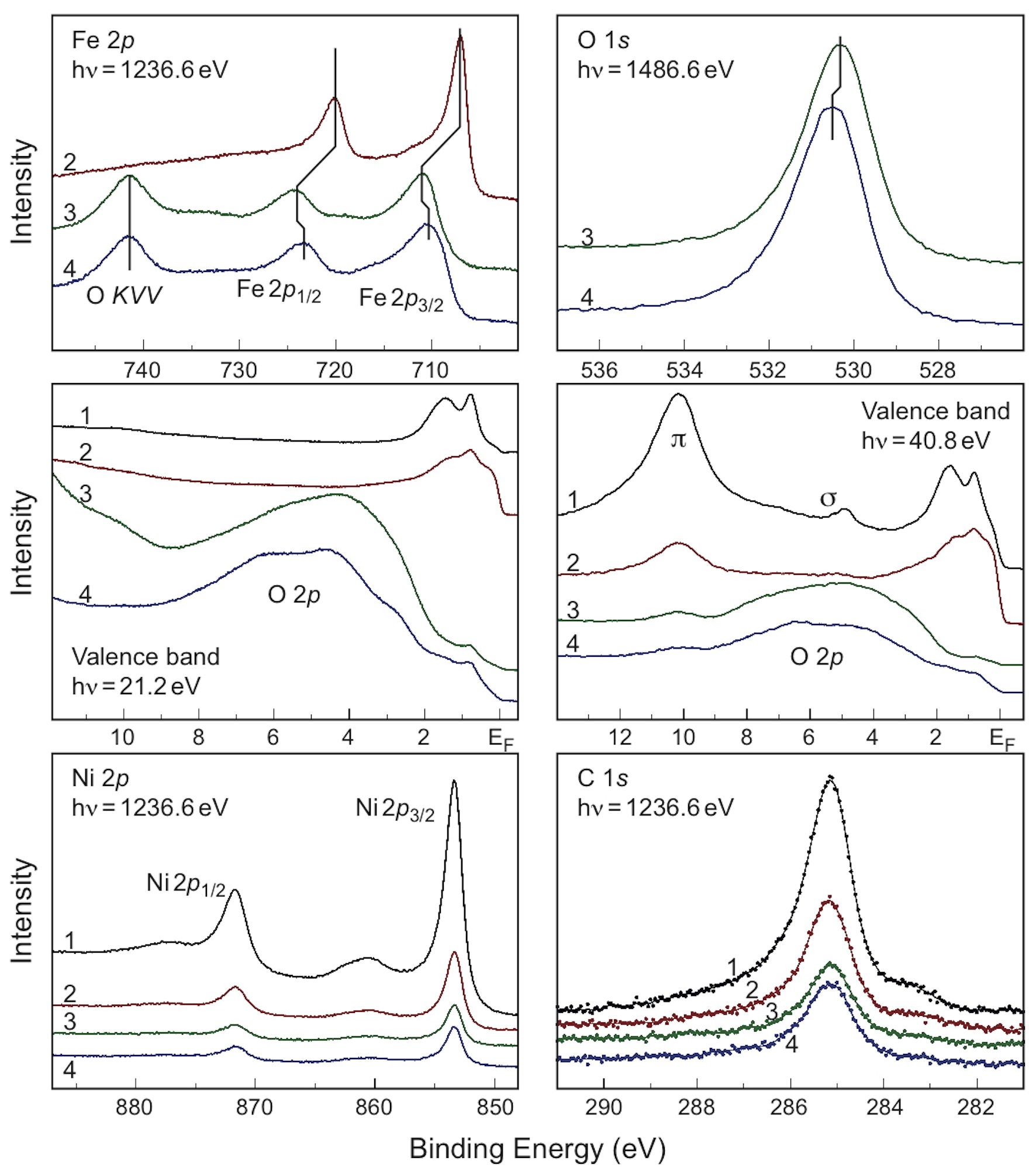}
\newline
\newline

\caption{(Color online).}
\label{xps}
\end{figure}

Figure~\ref{xps} shows XPS core-level spectra and PES spectra of the valence-band region obtained during preparation of the Fe$_3$O$_4$(111)/graphene/Ni(111) trilayer (the corresponding signs for core-levels and valence-band regions as well as respective photon energies are marked in the figure). The spectra are numbered by corresponding preparation steps: 1\,--\,graphene/Ni(111); 2\,--\,10\,\AA\,Fe/graphene/Ni(111); 3\,--\,O$_2$/10\,\AA\,Fe/graphene/Ni(111); 4\,--\,Fe$_3$O$_4$(111)/graphene/Ni(111). Deposition of Fe on graphene/Ni(111) (spectra 1 and 2 in Fig.~\ref{xps}) yields the decreasing of the photoemission intensities of Ni\,$2p$, C\,$1s$, and graphene\,$\pi$ emission lines leaving their binding energy positions unchanged. At the same time, valence-band region in the vicinity of the Fermi level is strongly modified due to the overlapping of emission features of Ni\,$3d$ and Fe\,$3d$ valence band states. 

Room-temperature oxidation of the 10\,\AA\,Fe/graphene/Ni(111) system leads to the strong modification of the XPS Fe\,$2p$ core-level and PES valence-band spectra (spectra 3 in Fig.~\ref{xps}). Chemical shift to higher binding energies of the Fe\,$2p$ emission lines (with absence of any metallic contribution) and the appearance of the strong O\,$2p$ emission in the valence band region indicate formation of iron oxide film with the highest Fe$^{3+}$ oxidation state, Fe$_2$O$_3$. Binding energy of the Fe\,$2p_{3/2}$ emission line is $711.0\pm0.1$\,eV (also its shape with characteristic satellite peaks at $719.0$\,eV and $733.1$\,eV) and the shape of the valence-band spectra measured with $21.2$\,eV and $40.8$\,eV are consistent with the recently published results for this compound~\cite{Xue:2011}. At the same time the graphene layer on Ni(111) behaves as a protection layer preventing any oxidation of the ferromagnetic Ni film that can be deduced from the XPS Ni\,$2p$ and C\,$1s$ spectra. These spectra show only decreasing in intensity indicating an increasing of the thickness of the iron oxide overlayer compared to the system before oxidation. Graphene\,$\pi$ states show the same behavior: binding energy is unchanged and intensity is strongly suppressed. 

Brief annealing of the oxidized iron layer on graphene/Ni(111) leads to the dramatic changes in the photoemission spectra, core-level as well as valence-band, that can be taken as an indication of the structural and electronic transformation of the iron oxide film (LEED spots in Fig.~\ref{LEED}(c) and spectra 4 in Fig.~\ref{xps}). Firstly, this modification is reflected in the energy shift of the Fe\,$2p$ emission line to smaller binding energies which are characteristic for the mix-valent iron oxide, magnetite, Fe$_3$O$_4$. The corresponding energy of Fe\,$2p_{3/2}$ is $710.4\pm0.1$\,eV. This value is consistent with respective energies of this line as reported in literature for Fe$_3$O$_4$~\cite{Xue:2011,Weiss:2002}. Satellite peaks visible in the Fe\,$2p$ XPS spectra of iron oxide before annealing (spectrum 3 for Fe$_2$O$_3$) are also not recognizable in this spectrum (spectrum 4). The corresponding shift is also detected for O\,$1s$ core level but in opposite direction as expected. The energy shift of this level is $150$\,meV as deduced from XPS data. The respective changes are also observed in the photoemission spectra of the valence band of the oxide layer measured with $21.2$\,eV and $40.8$\,eV of photon energy. Photoemission feature at $0.83$\,eV, which reflects emission from Fe$^{2+}\,3d^6$ initial state, as well as intense emission from the O\,$2p$ valence-band states between 2\,eV and 8\,eV of binding energy become more structured and these spectra are similar to those reported in the literature for Fe$_3$O$_4$~\cite{Dedkov:2004a,Xue:2011}. At the same time intensity of the graphene\,$\pi$ emission is slightly reduced indicating the formation of smooth layer of iron oxide Fe$_3$O$_4$. XPS Ni\,$2p$ as well as C\,$1s$ spectra are not changed during this annealing procedure confirming formation of the sharp interface system Fe$_3$O$_4$(111)/graphene/Ni(111) where electronic structure of the graphene layer on Ni(111) is not affected by the deposited iron oxide layer.

\begin{figure}[t]
\includegraphics[scale=1.0]{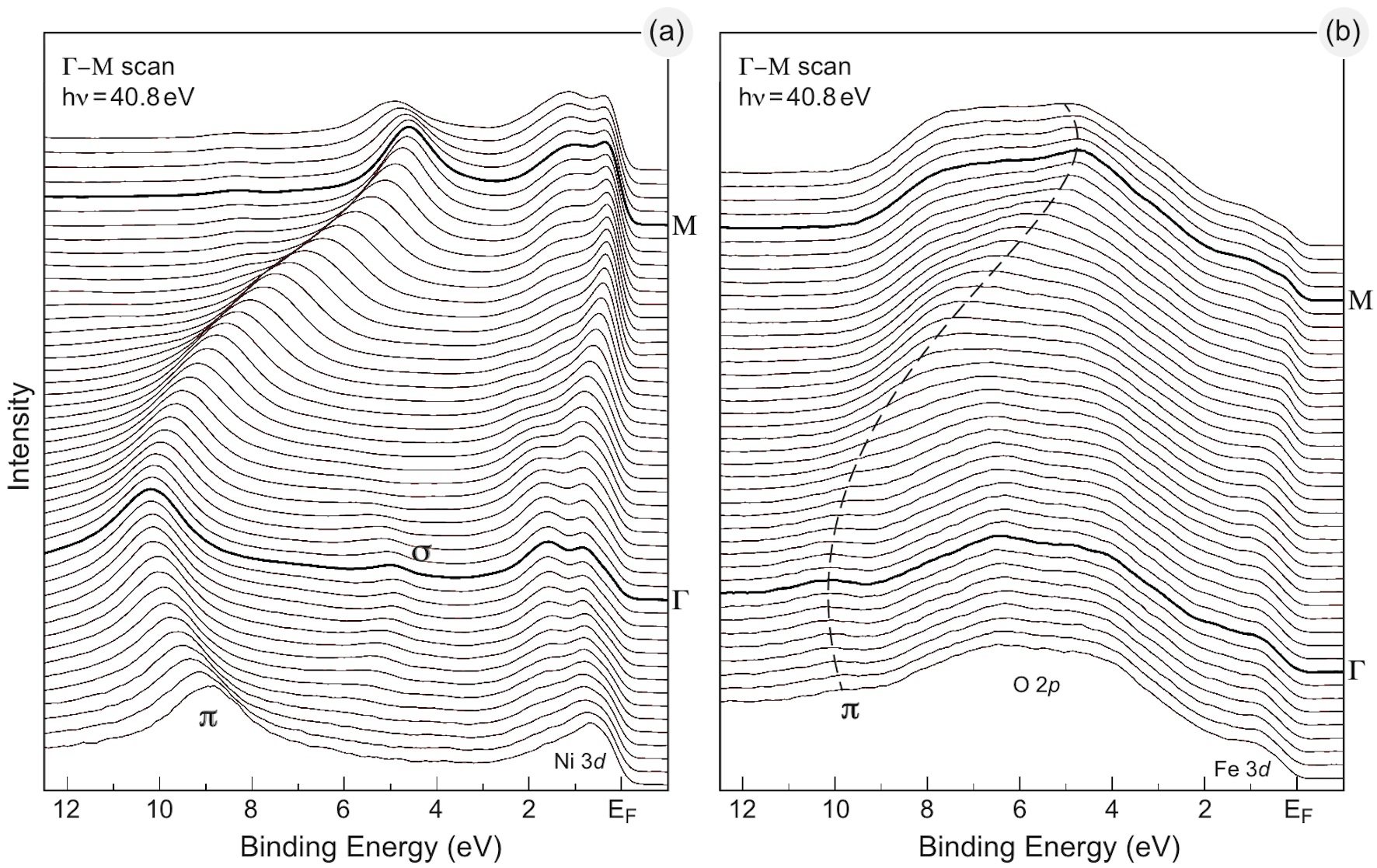}
\newline
\newline
\caption{(Color online).}
\label{arpes}
\end{figure}

The high quality of the formed trilayer was confirmed by angle-resolved PES (ARPES) measurements of its valence band (Fig.~\ref{arpes}). This figure shows stacks of ARPES spectra measured with $40.8$\,eV of photon energy along $\Gamma-M$ direction of the graphene Brillouin zone for (a) the graphene/Ni(111) and (b) the Fe$_3$O$_4$(111)/graphene/Ni(111) systems. ARPES spectra of graphene/Ni(111) [Fig.~\ref{arpes}(a)] are in very good agreement with previously published data and are described in the framework of the strong hybridization of the graphene\,$\pi$ and Ni\,$3d$ valence band states~\cite{Dedkov:2008a,Dedkov:2010a,Dedkov:2001}. As discussed in the previous paragraph, formation of thin Fe$_3$O$_4$(111) layer on graphene/Ni(111) leads to the suppression of intensities of the graphene- and Ni-related features but they still can be traced in ARPES spectra. For example, the dispersion of the graphene $\pi$ states is guided by the dashed line in Fig.~\ref{arpes}(b). The main emission features in these spectra are Fe\,$3d$ and O\,$2p$ in the range of $0-2$\,eV and $2.5-9$\,eV of binding energy, respectively. They are almost dispersionless contrary to results presented in Ref.~\cite{Dedkov:2004a}, that can be assigned to the most preferential growth of magnetite film on W(110) in the former case compared to the present situation where the lattice mismatch between graphene and magnetite lattices is quite large. However, this fact does not destroy the idea of the ideal spin-filtering effect in the discussed trilayer system because in any case one spin channel (spin-down) is not available for electrons which enter into Fe$_3$O$_4$ drain.

\textit{In conclusion}, the graphene-based spin-filter is proposed where half-metallic magnetite film is used as a detector of spin-polarized electrons which filter out one of the spin-channels. Fe$_3$O$_4$(111)/graphene/Ni(111) trilayer was prepared and characterized by means of LEED and PES of core levels and the valence band of the system. These results demonstrate that sharp interfaces between graphene and ferromagnetic films are formed in this system and electronic structure of the graphene/Ni(111) stays almost intact upon formation of thin magnetite layer on top. We suggest that such system helps to overcome major difficulties in realization of the graphene-based spin-filter which are discussed in the literature.
 
This work has been supported by the European Science Foundation (ESF) under the EUROCORES Programme EuroGRAPHENE (Project ``SpinGraph''). Y.\,D. acknowledges the financial support by the German Research Foundation under project DE\,1679/2-1.


\newpage

\textbf{Figure captions:}
\\
\\
\textbf{Fig.\,1.}\,(Color online) (a,b) Principles of graphene-based spin-filters where spin-polarized electrons are injected from ferromagnetic contact in graphene and detected by the second ordinary or half-metallic ferromagnetic contact. In the second case one of the spin channels (spin-down) is filtered out. Crystallographic structures of (c) the graphene/Ni(111) system in the $top-fcc$ configuration and (d) the Fe$_3$O$_4$(111) surface. Small solid-line hexagons, dashed-line hexagon, and big solid-line hexagon mark graphene unit cell, pseudo ``$1\times1$'', and unit cell of Fe$_3$O$_4$(111), respectively. 
\\
\\
\textbf{Fig.\,2.}\,(Color online) LEED images of (a) graphene/Ni(111), (b) the system obtained after oxidation of 10\,\AA\ Fe on graphene/Ni(111), and systems obtained after annealing of the later at (c) $150^\circ$C and (d) $250^\circ$C, respectively. Bright arrows indicate reflexes corresponding to the $(2\times2)$ structure of the Fe$_3$O$_4$(111) surface. 
\\
\\
\textbf{Fig.\,3.}\,(Color online) XPS core-levels and PES valence-band spectra (corresponding notations and photon energies are marked in the figure) of 1\,--\,graphene/Ni(111), 2\,--\,10\,\AA\,Fe/graphene/Ni(111), 3\,--\,O$_2$/10\,\AA\,Fe/graphene/Ni(111), and 4\,--\,Fe$_3$O$_4$(111)/graphene/Ni(111). For details of preparation, see text.
\\
\\
\textbf{Fig.\,4.}\,(Color online) Series of ARPES spectra measured with $40.8$\,eV of photon energy on (a) the graphene/Ni(111) system and (b) the Fe$_3$O$_4$(111)/graphene/Ni(111) trilayer. The energy dispersion of the graphene $\pi$ states is guided by the dashed line in (b).

\newpage

\newpage

\newpage

\newpage

\end{document}